\documentclass{mn2e} \usepackage{graphics} \usepackage{epsfig}

\def\ni{\noindent}                                    
\def\etal{et\thinspace al.\thinspace}                    

\title[Uncovering the chemical enrichment and mass-assembly histories of star-forming galaxies]
      {Uncovering the chemical enrichment and mass-assembly histories of star-forming galaxies}

\author[Cid Fernandes et al]
         {R. Cid Fernandes$^{1}$,
	  N. V. Asari$^{1}$,
	  L. Sodr\'e Jr.$^{2}$,
	  G. Stasi\'nska $^{3}$,
 	  A. Mateus$^{2}$, 
	 \newauthor
	 J. P. Torres-Papaqui$^{1}$,
	 W. Schoenell $^{1}$ \\
	 $^{1}$Departamento de F\'{\i}sica - CFM - Universidade Federal de Santa Catarina,
	 Florian\'opolis, SC, Brazil\\
	 $^{2}$Instituto de Astronomia, Geof\'{\i}sica e Ci\^encias
         Atmosf\'ericas, Universidade de S\~ao Paulo, S\~ao Paulo, SP,
         Brazil\\
	 $^{3}$LUTH, Observatoire de Meudon, 92195 Meudon Cedex, France}

\begin{document}

\maketitle

\begin{abstract} 
We explore the mass-assembly and chemical enrichment histories of star
forming galaxies by applying a population synthesis method to a sample
of 84828 galaxies from the Sloan Digital Sky Survey Data Release
5. Our method decomposes the entire observed spectrum in terms of a
sum of simple stellar populations spanning a wide range of ages and
metallicities, thus allowing the reconstruction of galaxy histories.
A comparative study of galaxy evolution is presented, where galaxies
are grouped onto bins of nebular abundances or mass. We find that
galaxies whose warm interstellar medium is poor in heavy elements are
slow in forming stars. Their stellar metallicities also rise slowly
with time, reaching their current values ($Z_\star \sim 1/3 Z_\odot$)
in the last $\sim 100$ Myr of evolution. Systems with metal rich
nebulae, on the other hand, assembled most of their mass and completed
their chemical evolution long ago, reaching $Z_\star \sim Z_\odot$
already at lookback times of several Gyr. These same trends, which
are ultimately a consequence of galaxy downsizing, appear when
galaxies are grouped according to their stellar mass. The
reconstruction of galaxy histories to this level of detail out of
integrated spectra offers promising prospects in the field of galaxy
evolution theories.
\end{abstract}


\section{Introduction}
\label{sec:Introduction}

One of the major challenges of modern astrophysics is to understand
the physical processes involved in galaxy formation and
evolution. Significant steps in this direction could be made by
tracing the build up of stellar mass and metallicity as a function of
cosmic time.

One way to address this issue is through cosmologically deep surveys
which map how galaxy properties change for samples at different
redshifts ($z$). Among these properties, the relation first observed by
Lequeux \etal (1979) between heavy-element nebular abundance and
galaxy mass, or its extension, the luminosity-metallicity relation
(e.g., Skillman \etal 1989; Zaritsky \etal 1994), are being
extensively used to probe the metal enrichment along cosmic
history. Clear signs of evolution are being revealed by studies of
these relations at both intermediate (Savaglio \etal 2005; Lamareille
\etal 2006; Mouhcine \etal 2006) and high $z$ (Shapley \etal 2005;
Maier \etal 2006; Erb \etal 2006), which generally find significant
offsets in these relations when compared to their versions in the
local Universe.

Alternatively, one may study galaxy evolution through {\em fossil}
methods, which reconstruct the star formation history (SFH) of a
galaxy from the properties of their stars. The most detailed studies
of this sort are those based on nearby galaxies (including our own),
where color-magnitude diagrams and even spectroscopy of individual
stars are feasible (e.g., Smecker-Hane \etal 1996; Hernandez \etal
2000; Dolphin 2002; Rizzi \etal 2003; Skillman \etal 2003; Monaco
\etal 2005; Koch \etal 2005). Comparing such observations with
predictions of evolutionary synthesis models then allows one to infer
how stars were born as a function of time, as well as to constrain the
chemical enrichment histories of the galaxies (e.g., Lanfranchi,
Matteucci \& Cescutti 2006; Carigi, Col\'in \& Peimbert 2006). By
mapping how stars become richer in metals as a galaxy ages, these
studies provide useful constraints for chemical evolution models.

Beyond the local group and its neighborhood, fossil methods must
derive SFHs from integrated galaxy spectra, using either selected
spectral features (Kauffmann \etal 2004; Brinchmann \etal 2004;
Gallazzi \etal 2005; Nelan \etal 2005; Thomas \etal 2005; Bernardi
\etal 2006) or the full spectrum (Panter, Heavens \& Jimenez 2003;
Heavens \etal 2004; Cid Fernandes \etal 2005; Mathis, Charlot \&
Brinchmann 2006; Panter \etal 2006).  This approach works with much
less information than is available from individual stars in galaxies,
so one cannot expect similar levels of accuracy. On the other hand,
integrated spectra of galaxies are now available by the hundreds of
thousands from modern spectroscopic surveys, so the decrease in
accuracy is compensated by orders of magnitude increase in statistics
and by the broader range in galaxy types spanned by such surveys. It
is therefore of great interest to explore methods to retrieve
information on galaxy evolution from integrated spectroscopic data.

In our ongoing series of papers entitled the Semi Empirical Analysis
of Sloan Digital Sky Survey (SDSS) Galaxies, (Cid Fernandes \etal
2005; Mateus \etal 2006a; Stasi\'nska \etal 2006; Mateus \etal 2006b;
SEAGal I--IV, respectively) we have shown that a decomposition of
galaxy spectra in terms of instantaneous bursts of different ages
($t_\star$) and metallicities ($Z_\star$) produces both excellent
spectral fits and astrophysically meaningful results. So far, however,
our description of SFHs has been extremely sketchy, based on averages
over the $t_\star$ and $Z_\star$ distributions, thus washing out
valuable time-dependent information.

This paper shows that we can actually do much better.  To illustrate
this we have culled a sample of star-forming (SF) galaxies from the
SDSS (Section \ref{sec:Data}), to which we apply our synthesis method
and derive mass-assembly and chemical enrichment histories (Section
\ref{sec:SFH_theory}). The evolution of stellar mass and metallicity
is then studied by grouping galaxies with similar nebular abundances
or mass (Section \ref{sec:SFH_results}). Details of this analysis and
cross-checks on the results reported here are presented in a companion
paper (Asari \etal, in prep, hereafter A07).

\section{Data}
\label{sec:Data}

\begin{figure*}
\includegraphics[width=0.9\textwidth]{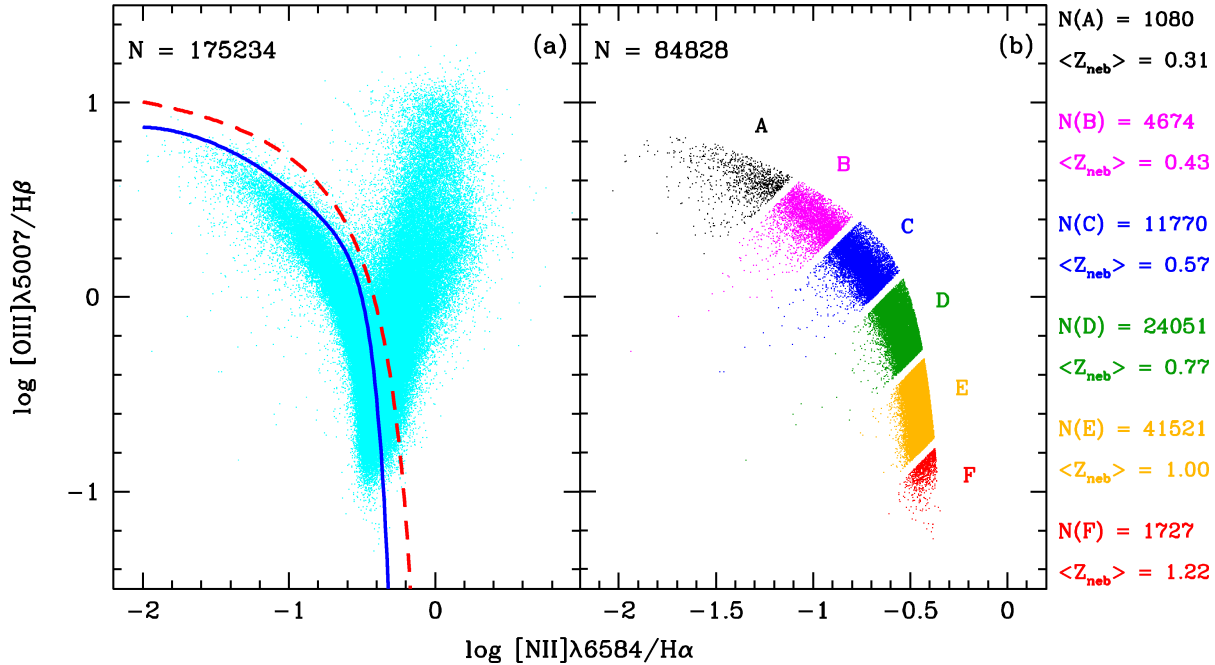}
\caption{(a) 175234 SDSS galaxies in the BPT diagram. The solid
line (Stasi\'nska et al 2006) divides Star-Forming galaxies from those
hosting AGN. The dashed line shows the dividing line used by Kauffmann
\etal (2003).  (b) The SF sample, chopped into six bins of nebular
abundance. The number of galaxies in each bin is given in the right,
along with the corresponding mean $Z_{neb}$ values (in solar
units). Galaxies close to bin borders are not plotted for clarity.}
\label{fig:BPT}
\end{figure*}

We have recently completed detailed pixel-by-pixel spectral fits of
573141 galaxies from the Main Galaxy Sample of the SDSS Data Release 5
(Adelman-McCarthy \etal 2006) with the synthesis code STARLIGHT
described in SEAGal I and II. STARLIGHT decomposes an observed
spectrum in terms of a sum of simple stellar populations covering a
grid of 25 ages from $t_{\star,j} = 1$ Myr to 18 Gyr and 6
metallicities $Z_{\star,j} = 0.0001$ to 0.05 (ie, from 0.005 to 2.5
$Z_\odot$). Each of these $N_\star = 150$ instantaneous bursts is
represented by a spectrum extracted from the evolutionary synthesis
models of Bruzual \& Charlot (2003) for a Chabrier (2003) initial mass
function, Padova 1994 tracks, and STELIB library (Le Borgne \etal
2003). Illustrative fits are presented in SEAGal I and A07.

Emission lines were measured from the starlight subtracted spectra.
Figure \ref{fig:BPT}a shows the Baldwin, Phillips \& Terlevich (1981,
BPT) diagnostic diagram for 175234 galaxies for which H$\beta$,
[OIII]$\lambda$5007, H$\alpha$ and [NII]$\lambda$6584 are all detected
with $S/N \ge 3$.  Reddening corrected line fluxes are used in this
plot and throughout our analysis. The distribution of points in this
diagram resembles a flying seagull, with SF galaxies occupying the
left wing and galaxies hosting an AGN on the right. We define SF
galaxies as those under the divisory line proposed in SEAGal III
(solid line in Figure \ref{fig:BPT}a). Further requiring a minimum
continuum $S/N$ of 10 at $\sim 4750$ \AA\ leaves a set of 84828
objects, hereafter the SF-sample, shown in Figure \ref{fig:BPT}b.
Results for alternative sample definitions (involving different $S/N$
thresholds, other AGN/SF separation criteria, and cuts on galaxy
inclination) are presented in A07. Note that by selecting SF galaxies
we indirectly select spirals over elliptical galaxies.

The SF-wing is essentially a sequence in nebular metallicity (SEAGal
III and references therein), which we quantify by the oxygen abundance
obtained through the strong-line calibration of Stasi\'nska (2006):

\begin{equation}
\label{eq:Z_neb}
\log Z_{neb} = 
   \log \frac{{\rm (O/H)}}{{\rm ~~(O/H)}_\odot} = 
   -0.14 - 0.25 \log \frac{{\rm [OIII]}\lambda5007}{{\rm [NII]}\lambda6584}
\end{equation}

\noindent where ${\rm (O/H)}_\odot = 4.9 \times 10^{-4}$ (Allende
Prieto, Lambert \& Asplund 2001). We believe this is the most reliable
calibration available in the literature, since it is based on
temperature-based abundances in the metal-poor case and on the ([OII]
+ [OIII])/H$\beta$ calibration of Pilyugin (2000) in the metal-rich
regime, using a large sample of giant HII regions.

$Z_{neb}$ is both a physically motivated and mathematically convenient
coordinate to map galaxy positions along the left wing in the BPT
diagram. From the tip of the SF-wing in Fig \ref{fig:BPT}a to its
bottom, our SF sample spans the $Z_{neb} \sim 0.2$--1.6 $Z_\odot$
range. In Fig \ref{fig:BPT}b this interval is chopped onto 6
logarithmic bins of width $\Delta \log Z_{neb} = 0.13$ dex, except for
the first one which is twice as wide to include more sources.
Galaxies inside these same bins will be grouped together in the
analysis of star-formation and chemical enrichment histories presented
in the following.

\section{Method of analysis}
\label{sec:SFH_theory}

As thoroughly discussed in the literature on stellar population
synthesis, the combination of noise in the data, astrophysical and
mathematical degeneracies limits the amount of information about
galactic histories that can be extracted from integrated
spectra. Clearly, the 150 components of the population vector used in
our fits are a greatly over-detailed description, so some {\em a
posteriori} compression of the results to a lower dimension is
needed. Our previous papers (SEAGal I--IV) have taken this compression
approach to its extreme by reducing the whole $t_\star$ and $Z_\star$
distributions encoded in the population vector to their first moments,

\begin{equation}
\langle \log t_\star \rangle_M = 
   \sum_{j=1}^{N_\star} \mu_j \log t_{\star,j}
\end{equation}

\begin{equation}
\langle Z_\star \rangle_M = 
   \sum_{j=1}^{N_\star} \mu_j Z_{\star,j}
\end{equation}

\ni where $\mu_j$ is the fraction of the total stellar mass
($M_\star$) which is in the $j^{th}$ population. 

Both $\langle \log t_\star \rangle$ and $\langle Z_\star \rangle$
proved to be mathematically robust and astrophysically valuable
summaries of the full population vector. In SEAGal I, II and A07 these
averages are shown to cleanly reveal important relations, such as
those between $\langle \log t_\star \rangle$ and $M_\star$ (associated
with the galaxy downsizing phenomenon), $\langle Z_\star \rangle$ and
$M_\star$ (the mass-metallicity relation), and $\langle Z_\star
\rangle$ and $Z_{neb}$ (which reveals a connection between stellar and
nebular chemical enrichment levels).  Given this success, it is
natural to explore less drastic compressions of the population vector.

One way to compress the population vector but retain the
age-information is to marginalize over the $Z_{\star,j}$ distribution
and compute the total mass converted into stars as a function of time:

\begin{equation}
\label{eq:MAH}
\eta^c_\star(t_\star) = 
    \sum_{t_{\star,j} > t_\star} \mu^c_j
\end{equation}

\ni where $\mu^c_j$ denotes the fraction of the total mass converted
onto stars ($M^c_\star$) that is associated with the $j^{th}$ base
age. The superscript ``c'' stands for ``converted onto stars'', and is
introduced to distinguish $\mu^c_j$ and $M_\star^c$ from $\mu_j$ and
$M_\star$, which refer to the mass still locked inside stars, ie,
discounting the mass returned to the inter-stellar medium (ISM) by
stellar evolution. $\eta^c_\star(t_\star)$ is thus a monotonic
function which grows from 0 to 1 as $t_\star$ runs from the time
galaxy formation starts to the present, and maps what fraction of
$M^c_\star$ was assembled up to a given lookback time.\footnote{Since
our main goal is to compare the {\em intrinsic} evolution of galaxies,
throughout this letter we consider ages in the context of
stellar-evolution alone, ie., as if all galaxies were observed at $z =
0$, such that $t_\star$ is both a stellar-age and a lookback time.}

To track the evolution of the {\em stellar metallicity} we compute the
total mass in metals inside stars as a function of $t_\star$ and
divide it by the total mass inside stars at the same time. This yields
the time-dependent mean stellar metallicity:

\begin{equation}
\label{eq:zeta}
\overline{Z_\star}(t_\star) = 
\sum_{t_{\star,j} > t_\star} \mu_j(t_\star) Z_{\star,j}
\end{equation}

\ni The computation of $\mu_j(t_\star)$ takes into account that the
total mass inside stars changes with time, both due to the SFH and to
the time dependence of the returned mass. Accordingly, only
populations older than $t_\star$ enter the definition of
$\mu_j(t_\star)$, since younger ones had not formed yet, ie.,
$\mu_j(t_\star) = 0$ for $t_{\star,j} < t_\star$.

\section{Results}
\label{sec:SFH_results}

The explicit time-dependence in equations (\ref{eq:MAH}) and
(\ref{eq:zeta}) provides a description of the star formation and
chemical histories of galaxies which goes well beyond that obtained
with the mere first moments of the age and metallicity
distributions. Given the numerous degeneracies which affect population
synthesis of galaxy spectra one might be skeptical as to how useful
they actually are. Simulations addressing this issue will be presented
elsewhere. In what follows we report what is obtained in practice.

\subsection{Galaxy evolution in $Z_{neb}$-bins}

Our strategy is to chop the SF-wing into bins in $Z_{neb}$ and study
the evolution of galaxies from the statistics of
$\eta^c_\star(t_\star)$ and $\overline{Z_\star}(t_\star)$ in each
bin. The fact that $Z_{neb}$ correlates strongly with several physical
and observational properties guarantees that galaxies inside each
$Z_{neb}$ bin are intrinsically similar, which in turn implies that
statistics over such bins is meaningful.

\begin{figure}
  \includegraphics[width=0.5\textwidth]{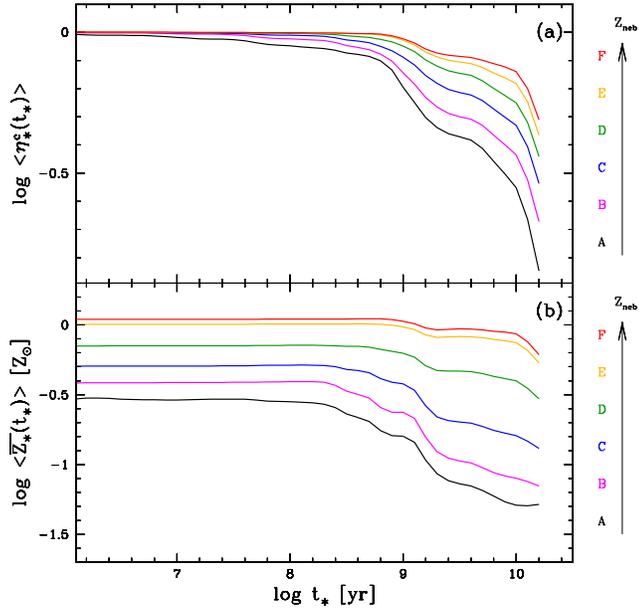}
\caption{(a) Average mass assembly ($\eta^c_\star$) and (b) chemical
evolution ($\overline{Z_\star}$) of stars in SF galaxies for the same
six $Z_{neb}$-bins defined in Fig \ref{fig:BPT}b. Each line represents
a $t_\star$-by-$t_\star$ average over all galaxies in the bin.  A mild
gaussian smoothing with FWHM = 0.2 dex in $\log t_\star$ was applied
to the mean $\eta^c_\star(t_\star)$ and $\overline{Z_\star}(t_\star)$
curves, just enough to smooth out discreteness effects associated to
the 25 ages in the base.}
\label{fig:mean_ChemEvol}
\end{figure}

Fig \ref{fig:mean_ChemEvol}a shows the mean mass-assembly functions
obtained for each of the six $Z_{neb}$ bins illustrated in Fig
\ref{fig:BPT}b. The systematics is evident to the eye. Galaxies with
low $Z_{neb}$ are slower in assembling their stars than those with
high $Z_{neb}$.  This behavior reflects the fact that, although most
of the mass is assembled at early times ($t_\star \ga 1$ Gyr),
galaxies at the tip of the SF-wing have larger star formation rates
(SFR) per unit mass in the recent past than those at the bottom of the
wing. This trend can be quantified in terms the ratio of the mean SFR
in the last 60 Myr to that over the galaxy's history, ie, the so
called Scalo's $b$ parameter (eg, Brinchmann \etal 2004). We find $b$
to decrease by a factor of $\sim 20$ from the lower to the higher
$Z_{neb}$ bins. (SFRs are discussed in A07, where it is also shown
that our synthesis based SFRs agree very well with those derived from
the H$\alpha$ luminosity.)  Alternatively, one may quantify this
behavior by computing the time at which the mean
$\eta^c_\star(t_\star)$ function reaches 0.75, ie., the time at which
the mass involved in star-formation reaches 3/4 of the total. We find
that $\tau_{75}$ increases from $\sim 0.8$ Gyr at $Z_{neb} \sim 0.31 $
(bin A in Fig \ref{fig:BPT}b), to 2 Gyr at $Z_{neb} \sim 0.77 Z_\odot$
(bin D), and 8 Gyr for the highest $Z_{neb}$ bin.  Given that
$Z_{neb}$ correlates strongly with $M_\star$ (Zaritsky \etal 2004;
Tremonti \etal 2004; Lee \etal 2006), this behavior is ultimately a
signature of galaxy downsizing (Cowie \etal 1996).

The chemical evolution of stars in SF galaxies is shown in Fig
\ref{fig:mean_ChemEvol}b. Again, the main message is clear: Galaxies
with a metal poor ISM are low are slow in their stellar chemical
enrichment. In the lower $Z_{neb}$ bins, $\overline{Z_\star}(t_\star)$
only reaches the current stellar metallicity level ($\sim 1/3
Z_\odot$) as recently as $t_\star \sim 100$ Myr. Galaxies with larger
$Z_{neb}$, on the other hand, have flatter and systematically higher
$\overline{Z_\star}(t_\star)$ curves.  These massive systems have
essentially completed their mass assembly and chemical evolution long
ago, so that recent star-forming activity has a negligible impact upon
$\overline{Z_\star}(t_\star)$.

The fact that bins of increasing $Z_{neb}$ have increasingly higher
present $\overline{Z_\star}$ values confirms that nebular and stellar
metallicities are related (SEAGal I; Gallazzi \etal 2005; A07). We
nevertheless warn the reader not to overinterpret the quantitative
relation between these two quantities, given that they are derived by
completely different and hardly comparable means.  In this paper we
use [OIII]/[NII] as an indicator of $Z_{neb}$ (equation
\ref{eq:Z_neb}), because of its evident relation to the position of
the points in the BPT diagram. However, as emphasized by Stasi\'nska
(2006), strong line methods for deriving nebular abundances have been
calibrated using giant HII regions and their relevance for integrated
spectra of galaxies has still to be examined, because of the role of
metallicity gradients and of the diffuse ionized medium in
galaxies. In addition, there is so far no study comparing nebular
abundance calibrations with stellar abundances. Therefore, our
$Z_{neb}$ values are not necessarily on the same scale as stellar
abundances. Insofar as this paper is concerned, this is not a problem,
since, irrespective of absolute values, it is the existence of a
$Z_{neb}$-$\overline{Z_\star}$ relation which is responsible for the
ordering of the different chemical evolution curves in Fig
\ref{fig:mean_ChemEvol}b.

\begin{figure}
  \includegraphics[width=0.5\textwidth]{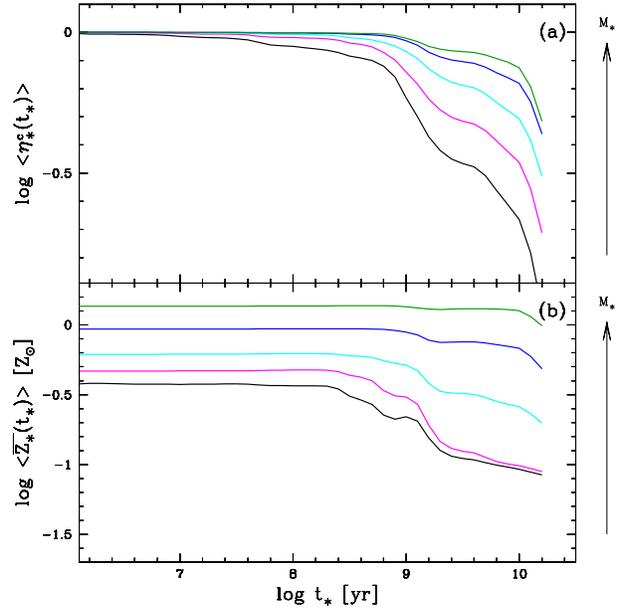}
\caption{As Fig \ref{fig:mean_ChemEvol}, but binning SF galaxies by
their stellar mass, using five 1 dex wide bins, centered at (from
bottom to top) $\log M_\star/M_\odot = 7.5$, 8.5, 9.5, 10.5 and 11.5,
which contain 583, 4750, 29056, 46847 and 3550 galaxies,
respectively.}
\label{fig:mean_ChemEvol_BinnedByMass}
\end{figure}

\subsection{Galaxy evolution in Mass-bins}

While Fig \ref{fig:mean_ChemEvol} illustrates the strong link between
the stellar mass assembly and chemical evolution histories of galaxies
and the present ionized ISM metal content, $Z_{neb}$ is not the cause,
but the product of galaxy evolution, so its use as an independent
variable, although pedagogically useful, is questionable. If one had
to chose a single quantity as the main driver of galaxy evolution this
would surely be its mass (eg., Tinsley 1968; Larson 1974; Pagel 1997).

We have thus repeated the analysis above binning galaxies by $M_\star$
instead of $Z_{neb}$. Results are shown in Fig
\ref{fig:mean_ChemEvol_BinnedByMass} for five 1 dex-wide mass bins
centered at $\log M_\star/M_\odot = 7.5 \ldots 11.5$.  Because of the
flattening of the $M_\star-Z_{neb}$ relation at high masses and the
substantial scatter around the relation, particularly at low $M_\star$
and $Z_{neb}$, binning in $M_\star$ or $Z_{neb}$ samples somewhat
different populations of galaxies.  On the whole, however, this
alternative grouping scheme leads to the same general scenario
outlined above, with low $M_\star$ galaxies being slower in their mass
and chemical build up than more massive ones.  A caveat is that
$M_\star$ is not a perfect tracer of the depth of the potential well,
since it does not account for the mass in gas nor dark matter.  In any
case, the signature of galaxy downsizing is evident in Fig
\ref{fig:mean_ChemEvol_BinnedByMass}: massive galaxies today formed
most of their stars and metals long time ago, whereas low-mass
galaxies are still forming stars and metals actively.

The signal of chemical evolution is present in all curves in Figs
\ref{fig:mean_ChemEvol}b and \ref{fig:mean_ChemEvol_BinnedByMass}b.
Curiously, however, this evolution does not seem to start from very
low $\overline{Z_\star}$ for massive, high $Z_{neb}$ galaxies (eg,
bins E and F). We interpret this apparent lack of old, metal poor
populations as due to a combination of lack of age resolution at high
$t_\star$ and an intrinsically fast evolution ocurring at early
times. In low mass systems (eg, bin A), $\overline{Z_\star}(t_\star)$
starts is $\Delta t_\star \sim 2$ Gyr long rise period at $t_\star
\sim 2$--3 Gyr, such that $\Delta t_\star / t_\star$ is of order
unity. The same process happening at $t_\star > 10$ Gyr would yield
$\Delta t_\star / t_\star < 0.2$, and consequently be much harder to
resolve.  As shown above above, by $t_\star = 10$ Gyr massive galaxies
had already assembled most of their mass, so their chemical evolution
must also have occurred early on.  This mixture of populations of
widely different metallicities but similar ages results in
$\overline{Z_\star}$ values at $t_\star > 10$ Gyr which are dominated
by the metal rich stars already present at those times. Such an early
and rapid formation period should lead to enhancement of elements
synthesized in type II SNe (Worthey, Faber \& Gonz\'alez 1992; Thomas
\etal 2005). Though we cannot map elemental abundances with our
method, we do detect a signal of increasing $\alpha$ enhancement
towards the bottom of the BPT diagram by means of systematically
increasing residuals in spectral bands due to $\alpha$ elements like
Mg (A07), indirectly supporting the scenario outlined above.

\section{Concluding Remarks}

Our analysis shows that the present ISM abundance is intimately
connected to the past star-formation and chemical enrichment history
of a galaxy.  Whereas this is intuitively obvious from a physical
point of view, it is not obvious at all that one should be able to
recover this behavior from population synthesis analysis of integrated
galaxy spectra of SDSS-like quality. We emphasize that, unlike
previous studies such as that by Bica (1988), our fits impose no {\em
a priori} chemical evolution constraints whatsoever upon the mixture
of $t_{\star,j}$'s and $Z_{\star,j}$'s in the base. Given the absence
of constraints and the fact that the fits are completely independent
of the emission line data used to compute $Z_{neb}$, the organization
of galaxies with different $Z_{neb}$ onto the clear and systematically
distinct stellar chemical enrichment patterns seen in Fig
\ref{fig:mean_ChemEvol}b is remarkable, revealing that fossil methods
based on integrated spectra have reached a level of maturity which
would be unthinkable just a few years ago (eg, Cid Fernandes \etal
2001).  By taking the spectral synthesis output one step further than
the trivial first moments of the $t_\star$ and $Z_\star$
distributions, we have been able to uncover the chemical evolution of
galaxies with an unprecedented level of detail for such a large and
varied sample. This significant achievement was only possible due to
the fabulous statistics of the SDSS, combined with state-of-the-art
evolutionary synthesis models and reliable SFH recovery
techniques. The sheer number of galaxies allows us to recover an
unequivocal chemical evolution signal which would be at best doubtful
for individual galaxies or even small samples. Combined with our
estimates of the star-formation histories, the empirically derived
chemical evolution patterns should provide valuable constraints for
galaxy evolution models.

\section*{Acknowledgments}

We thank support from the Brazilian agencies CNPq and FAPESP. We are
greatly in debt with several colleagues and institutions around the
globe who have contributed to this project by allowing access to their
computers. The Sloan Digital Sky Survey is a joint project of The
University of Chicago, Fermilab, the Institute for Advanced Study, the
Japan Participation Group, the Johns Hopkins University, the Los
Alamos National Laboratory, the Max-Planck-Institute for Astronomy
(MPIA), the Max-Planck-Institute for Astrophysics (MPA), New Mexico
State University, Princeton University, the United States Naval
Observatory, and the University of Washington.  Funding for the
project has been provided by the Alfred P. Sloan Foundation, the
Participating Institutions, the National Aeronautics and Space
Administration, the National Science Foundation, the U.S. Department
of Energy, the Japanese Monbukagakusho, and the Max Planck Society.

\end{document}